\documentclass[10pt,twocolumn,prl,aps,floats,graphicx,epsf]{revtex4}
\usepackage{graphicx}
\def\a{\alpha}
\def\b{\beta}

\def\la{\langle}
\def\ra{\rangle}
\def\a{\alpha}

\def\b{\beta}

\def\h{\hskip 1cm}

\def\rl{\longleftrightarrow}

\begin{document}

\title
{ Perfect State Transfer in Two and Three Dimensional Structures}

\bigskip
\author{V. Karimipour$^1$\footnote{email:
vahid@sharif.edu}}
\author{M. Sarmadi Rad$^1$}
\author{M. Asoudeh$^{2}$}
\affiliation{$^{1}$Department of Physics, Sharif University of Technology, Tehran, Iran\\
$^2$ Department of Physics, Shahid Beheshti University, Tehran,
Iran }

\begin{abstract}
We introduce a scheme for perfect state transfer in regular two
and three dimensional structures. The interactions on the
lattices are of XX spin type with uniform couplings. In two
dimensions the structure is a hexagonal lattice and in three
dimensions it consists of hexagonal planes joined to each other
at arbitrary points. We will show that compared to other schemes,
i) much less control is needed for routing, ii) the algebra of
global control is quite simple, iii) the same kind of control can
upload and download qubit states to or from built-in Read-Write
(RW) heads. Pivotal use is made of a quantum switch which we
introduce and call it Hadamard switch, since its operation
depends on the existence of a Hadamard matrix in four dimensions.
\bigskip
\end{abstract}

\maketitle

\textbf{Introduction:} Since the inception of the fields of
quantum information and computation, the task of coherently
transferring quantum states, through long and short distances has
been of utmost importance. While photons are the ideal carriers
of quantum information over long distances, it has become evident
that the best possible method for transferring quantum
information over short distances, i.e. through regular arrays of
qubits, is to exploit the natural dynamics of the many body
system. This idea was first introduced in the work of Bose
\cite{bosePRL} who showed that the natural dynamics of a
Heisenberg ferromagnetic chain can achieve high-fidelity transfer
of qubits over distances as long as 80 lattice units.

In contrast to the traditional passive view in the study of many
body dynamics, where the dynamics is only calculated and observed,
this has led to an active view in the dynamics of solid state
systems, where we intervene in the dynamics to achieve certain
goals. For example it is tried to engineer the couplings in such
a way that states are transferred with perfect
\cite{Perfect,noninitialized} or with arbitrary high fidelity
\cite{Plenio,Wojcik,Bayat,Franco2,Feldman,Yao}, or it is tried to
interrupt the natural evolution by some minimal control to
achieve this task \cite{Burgarth1,Osborne,Cappellaro,Banchi}.

To overcome the necessity of engineered couplings, which severely
restricts experimental realization of such protocols, some kind
of control was re-introduced in the scheme \cite{Pemberton},
wherre it was shown in \cite{Pemberton} that quasi one
dimensional chains with uniform $\pm$ couplings, can achieve
perfect transfer. Conversion of these linear structures to star
configurations \cite{Pemberton} and arranging them in 2
dimensional structures was shown to achieve perfect state
transfer in two and higher dimensions. However the nature of
subsystems introduced in \cite{Pemberton} required multiple
control on external nodes of each subsystem, and different types
of control for matching subsystems with each other. A different
type of control was also necessary for loading and extracting the
states to or from the lattice.

In this letter we introduce a very simple scheme for perfect
transfer, in two and three dimensional lattices. In addition to
having all the nice properties of the protocol of
\cite{Pemberton}, like linear scaling of time with distance, and
robustness to errors, this scheme has very desirable extra
properties, namely:\\

\textbf{i)} very simple global operations are needed for routing
arbitrary states through arbitrary paths, that is, to each route
a very simple sequence of operations corresponds,\\

\textbf{ii)} the lattice has natural built-in local read-write
(RW) heads for uploading and downloading qubit states,\\

\textbf{iii)} the same kind of global control which is used for
routing, is also used for uploading and downloading states to or
from input and output heads.\\

As we will see, all these properties, are based on the geometry
of hexagonal lattice and on a concept (or device) which we
introduce for the first time. We call it the Hadamard switch,
since it uses the four dimensional Hadamard matrix.\\

In fact in \cite{Pemberton}, a localized (particle) state in one
of the input nodes, is considered as a superposition of complex
extended waves on all the input nodes where only one of these
waves passes through the star-shaped subsystem to the output
nodes and vice versa, while all the other waves are reflected
back. Therefore one needs to control the external nodes a few
times by applying suitable phase gates so that after multiple
traversal and reflections, the waves again interfere
constructively to create a particle state on a particular output
node. As we will show, the Hadamard switch and its natural
embedding in hexagonal lattices improves in a dramatic way all
these features. Moreover it naturally allows the imbedding of RW
heads in the lattice structure for uploading and downloading of
quantum states.

\textbf{Preliminaries} The prototype of many-body systems which
has been used in many protocols is the XY spin chain,
\begin{equation}\label{proto}
  H=\sum_{m,n} \frac{1}{2}K_{m,n}(X_mX_n+Y_mY_n).
\end{equation}
This type of interaction preserves spin $
  [H,\sum_m Z_m]=0$,
and moreover does not evolve the uniform background state of all
spin ups, i.e. $H^{\otimes N}|0\ra^{N}=0$, where $N$ is the total
number of qubits in the lattice. This then leads to the simple
result that for transferring an arbitrary qubit state like
$\a|0\ra+\b|1\ra$ it is enough to perfectly transfer only the
state $|1\ra$ through the lattice. Such a transfer occurs in the
single excitation sector which is spanned by $N$ states of the
form $|m\ra$, where $|m\ra$ means that the single spin in the
$m-$th place is down (or the local qubit is in the state
$|1\ra$). Note that we use the the names single-excitation and
single-particle states interchangeably, i.e. by the transfer of a
particle through the lattice, we mean transfer of excitation not
any actual particle.

It  was shown in \cite{Perfect} that a linear XY chain of length
N, with engineered local couplings in the form
$K_{n,n+1}=\sqrt{n(N-n)}$ can achieve perfect transfer between any
pair of input $n$ and output $N-n$ site in time $\pi$. It was
shown in \cite{Perfect} that the only uniformly coupled XY chains
which can do perfect transfer are of length 2 and 3 with
respective transfer times $t_0:= \frac{\pi}{2}$ and
$t_1:=\frac{\pi}{\sqrt{2}}$ (note that we have normalized the
coupling in both chains to unity).

\begin{figure}
\centerline{\includegraphics[scale=.16]{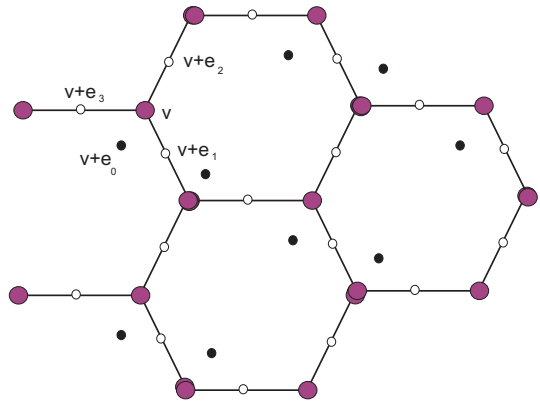}} \caption{(color
online) The hexagonal plane. The RW heads are are the small black
circles. The links accommodate qubit states (small white
circles). The Hadamard switches (the big circles at vertices) are
used to switch the qubit in different directions and the RW head
(if necessary).} \label{Tiling}
\end{figure}

\textbf{The scheme} We start with the hexagonal lattice shown in
figure (\ref{Tiling}). Let $v$ denote a vertex of the lattice. On
the three links connected to this vertex, there are three qubits,
which we denote by $v+e_1$, $v+e_2$ and $v+e_3$. The vectors
$e_1$, $e_2$ and $e_3$ denote the three vectors directed along
the links connected to a vertex. A fourth qubit $v+e_0$, called
the read-write (RW) head is also connected to this vertex,
although the vector $v+e_0$  does not necessarily mean a vector
in the plane and is used only for uniformity of notation. The
Hamiltonian which governs the interaction on this system is of
the form
\begin{equation}\label{Hsum}
  H = \sum_{v}H_v
\end{equation}
where $H_v$ is the local Hamiltonian connecting each vertex to
its neighboring links and through these links to the other
vertices.

\begin{figure}
\centerline{\includegraphics[scale=.11]{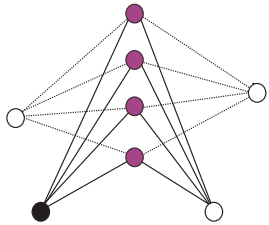}} \caption{(color
online) The Hadamard switch. The four spins in the middle (the
colored circles) have an XY interaction with the four spins
around according to the pattern of $\pm$ signs in the Hadamard
matrix $J$ in (\ref{Hadamard}). That there is no interaction
between the central spins. The white circles are the spins on the
link and the black one is the RW head.} \label{Joint}
\end{figure}

Now we describe the quantum system at each vertex. At each vertex
$v$ there are four qubits which we denote by $v_{_\a}$, i.e.
$v_{_0}, v_{_1}, v_{_2}, v_{_3}$. These particles are arranged on
four different layers so that all the qubits with the same index
lie in one layer. In particular the qubits $v_0$ for different
$v$'s lie in the hexagonal plane and the other qubits lie in
different layers which we call control layers to distinguish them
from the main hexagonal plane. As we will see later, the only
control that we need is the possibility of applying uniform
magnetic field on each control layer. No control on any
individual qubits is necessary. The interaction between all the
spins is a simple $XY$ interaction. Such an interaction conserves
the total z-component of spin and hence when the whole lattice is
initialized to the state $|0\ra$ (spin up), transfer of a
particle occurs in the single-particle sector. The local
Hamiltonian $H_v$ has the simple form
\begin{equation}\label{Hv}
  H_v = \sum_{\a,\b=0}^3 J^{\a\b} \left(X_{v_\a}X_{v+e_\b}+Y_{v_\a}Y_{v+e_\b}\right)
\end{equation}
where $J^{\a\b}$ are the entries of the Hadamard matrix in four
dimensions, namely
\begin{equation}\label{Hadamard}
  {J}=\frac{1}{2}\left(\begin{array}{cccc}1&1&1&1\\ 1&1&-1&-1\\ 1&-1&1&-1\\
  1&-1&-1&1\end{array}\right).
\end{equation}
We remind the reader that a Hadamard matrix is a symmetric real
orthogonal matrix all of whose entries have the same absolute
value. Such matrices exist only in certain special dimensions. We
will elaborate on the importance of this matrix for our scheme
later on. In the above matrix the rows and columns are numbered
from 0 to 3 from left to right and from top to bottom
respectively. Note that the vertex $v_0$ is connected with each of
the three links and also the RW head with equal couplings. We
call this structure, described by the Hamiltonian $H_v$ a
Hadamard switch. As we will see later, it can be used in a very
effective way for routing states through two and three
dimensional structures. Figure (\ref{Joint}) shows this switch.
In figure (\ref{Tiling}) these switches have been depicted as big
colored circles at vertices of the hexagonal lattice.

\begin{figure}
\centerline{\includegraphics[scale=.18]{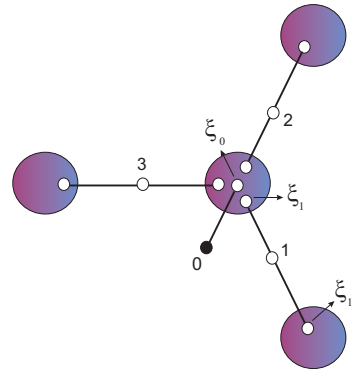}}
\caption{(color online) The effective $H_v$, when written in the
basis $|\xi^\a\ra$, decomposes into perfect transfer effective XY
chains of length two and three. The 2-chain transfers the particle
between the RW head (the black circle) and the state $|\xi^0\ra$.
Global control on the plane of spins 1, 2 or 3, changes the
states $|\xi^\a\ra$ as desired. A state like $|\xi^1\ra$
perfectly goes in the direction 1, while a state like $|\xi^2\ra$
is stopped in this direction and goes only through direction 2.
All the couplings on all the links are 1.} \label{BasicBranch2}
\end{figure}

It can be readily verified that the local XY hamiltonian, on
spins $m$ and $n$, $H_{m,n}:=\frac{1}{2}(X_mX_n+Y_mY_n)$ has the
following simple action on the single excitation states, $
H_{m,n}|m\ra=|n\ra, \ \ H_{m,n}|n\ra=|m\ra, $ where
$H_{m,n}|p\ra=0$ for $p\ne m, \ n$. This means that $H_{m,n}$
when restricted to single particle subspace has the following
expression,
\begin{equation}\label{Hmnket}
  H_{m,n}=|m\ra\la n|+|n\ra\la m|.
\end{equation}
This allows us to rewrite the total Hamiltonian in the form
\begin{equation}\label{Hvket}
H=\sum_{v,\a,\b}J^{\a,\b}(|v_\a\ra\la v+e_\b| + |v+e_\b\ra\la
v_\a|).
\end{equation}
We now consider the four\  orthogonal \ states \ $
  |\xi_v^{\a}\ra:=\sum_{\b=0}^3 J^{\a,\b}|v_\b\ra,
$
that is
\begin{eqnarray}\label{allxi}
  |\xi_v^{0}\ra&:=&\frac{1}{2}(|v_0\ra+|v_1\ra+|v_2\ra+|v_3\ra)\cr
 |\xi_v^{1}\ra&:=&\frac{1}{2}(|v_0\ra+|v_1\ra-|v_2\ra-|v_3\ra)\cr
  |\xi_v^{2}\ra&:=&\frac{1}{2}(|v_0\ra-|v_1\ra+|v_2\ra-|v_3\ra)\cr
   |\xi_v^{3}\ra&:=&\frac{1}{2}(|v_0\ra-|v_1\ra-|v_2\ra+|v_3\ra).
\end{eqnarray}
The Hamiltonian can now be rewritten as
\begin{equation}\label{Hxi}
H=\sum_{v,\a}(|\xi_v^\a\ra\la v+e_\a| + |v_\a\ra\la \xi_v^\a|).
\end{equation}
Thus as shown in figure (\ref{BasicBranch2}), in this new basis
the Hamiltonian has been decomposed into direct sum of XY spin
chains with uniform couplings of length two and three. We now
note that such chains are capable of perfect transfer of qubits
in times $t_0= \frac{\pi}{2}$ and $t_1=\frac{\pi}{\sqrt{2}}$
respectively.

It is important to note that the states $|\xi_v^{\a}\ra$ are
turned into each other by global unitary operators. Let $Z_1$,
$Z_2$ and $Z_3$ be the Pauli operators acting on spins 1, 2 and 3
on vertex $v$. Then it is readily seen that
\begin{eqnarray}\label{control}
Z^1Z^2:|\xi^1\ra\rl |\xi^2\ra,\h |\xi^0\ra\rl |\xi^3\ra,\cr
Z^1Z^3:|\xi^1\ra\rl |\xi^3\ra, \h |\xi^0\ra\rl |\xi^2\ra,\cr
Z^2Z^3:|\xi^2\ra\rl |\xi^3\ra,\h |\xi^0\ra\rl |\xi^1\ra.
\end{eqnarray}
The crucial point is that the $Z_i$ operations can be applied
globally on all the qubits in the $i-th$ control layer, since on
the empty sites it has no effect and on an occupied site it has
the phase effect that we want. Therefore there is no need for
addressing single spins in each control layer, only the
possibility of access to each layer is required. Such a control
should be applied in a time much shorter than the time scale of
evolution of the Hamiltonian, namely $t_0$ and $t_1$. If we want
to route many particles at the same time, we have to control
different regions of control layers in accordance with the paths
of these particles. The control will be the same in the time
intervals when the paths become parallel.

Now a clear and very simple method for perfect state transfer in
the lattice emerges. A single particle $\a|0\ra+\beta|1\ra$ is
uploaded to a given input head $v_{in}$. The part $\a|0\ra$ does
not evolve and indeed is ready for downloading at any output
head. We only have to transfer the single particle state $|1\ra$
which in view of our notation, has made the whole lattice to be
in the state $|v_{in}+e_0\ra$. After a time $t_0$, this state
evolves to $|\xi_{v_{in}}^0\ra$, i.e. the particle has moved to
the nearest vertex $v_{in}$ in the form of a real wave $\xi^0$.
Once in a state $|\xi_{v}^0\ra$, we can make a global control
according to (\ref{control}) to switch this vertex state to
either of the states $|\xi_v^i\ra$ ($i=1,2,3$) depending on the
direction we want to route the state. For example if we switch it
to $|\xi_{v}^1\ra$, then according to figure (\ref{BasicBranch2}),
after a time $t_1$, the state will be transferred perfectly to
the vertex $v+e_1$ in the form $|\xi_{v+e_1}^1\ra$. Continuing in
this way we can move the state via any path that we like to any
other vertex say $v_{out}$, where the final state will be one of
the three states $|\xi_{v_{out}}^i\ra$, ($i=1,2,3$). Switching
this state to $|\xi_{v_{out}}^0\ra$ will move this state to the
nearest output head in the form $|v_{out}+e_0\ra$ where it will
be read off. The total time for routing is $2t_0+Nt_1$, where $N$
is the number of links which connect the input and output heads
along the chosen path. The sequence of control operations is very
simple. For uploading and downloading a qubit to or from a link
$e_i$ to its nearest head, the operation $\hat{Z_i}$ is applied
when $\hat{Z_i}$ means that $Z_i$ is removed from the triple
$Z_1Z_2Z_3$ and at each vertex for turning the qubit from link
$v+e_i$ to link $v+e_j$, the operation $Z_iZ_j$ is applied.
Except for uploading and downloading operations where a time
lapse of $t_0$ is needed all the other control operations are
applied at regular intervals of
time $t_1$. We restate it as:\\

$\bullet:$ for turning from direction $i$ to $j$ apply $Z_iZ_j$,\\

$\bullet:$ For uploading and downloading a qubit to or from a RW
head to direction $i$ apply $\hat{Z_i}$.\\

\begin{figure}
\centerline{\includegraphics[scale=.18]{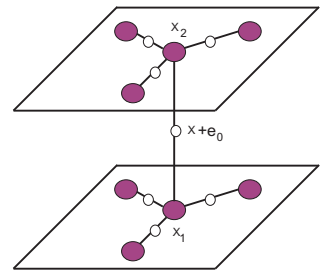}}
\caption{(color online) The Hadamard switch can transfer a state
between planes. The leg which was previously connected to RW
heads is used to connect two switches in two different planes.
The effective Hamiltonian on the 2-chain is now replaced by a
3-chain connecting two planes.} \label{twolayers}
\end{figure}

\textbf{Perfect transfer in three dimensions} The Hadamard switch
can be used in another way for achieving perfect state transfer
in three dimensional structures. Figure (\ref{Joint}) shows a
Hadamard switch connecting two hexagonal planes. Such planes can
be joined by any number of switches. The number and positions of
Hadamard switches are determined to optimize the accessibility of
all the heads in the two planes by shortest possible paths. When
used in this new way, the RW head gives its role to the qubit on
the link which joins the two planes. For example when the two
planes are joined to each other at points $x_1$, and $x_2$ on the
two planes (figure \ref{twolayers}), the effective Hamiltonian
for the states $|x+e_0\ra$ on the joining link and the states
$|\xi_{x_1}^{0}\ra $ on plane 1 and $|\xi_{x_2}^{0}\ra $ on the
upper plane is nothing but a perfect XY 3-chain. This effective
Hamiltonian, transforms the state $|\xi^0\ra$ perfectly between
the two planes. This time we should wait for time $t_1$ instead
of $t_0$. Therefore we route a particle within each plane as
before and bring it to the position of the nearest switch where by
appropriate control we move it to another plane and continue
there.

\textbf{Robustness} Like the system in \cite{Pemberton}, our
scheme is also robust against imperfections in the lattice up to
a threshold. That is if we know which of the switches are not
working we can easily route them around. Moreover due to the very
simple nature of the switches and their control, if there is any
delay in the control operations, we know exactly on which switch
and which link of the switch the particle is waiting. This is due
to the fact that under the intrinsic dynamics of the 2 and 3
chains in figure (\ref{BasicBranch2}), any excitation just goes
back and forth between the endpoints of a chain.

\textbf{Discussion} The Hadamard switch introduced in this paper
has a broader applicability than the one presented in this
scheme. In fact, given any type of perfect transfer chain
\cite{Perfect}, one can join them by Hadamard switches to each
other and to RW heads to form two and three dimensional
structures for perfect routing of quantum states. In this way one
can use fewer number of switches in each plane and also attach
different planes by longer chains which facilitates the access to
control planes. Compared to the start-structure of
\cite{Pemberton}, this is achieved at the cost of higher
connectivity in the switch. The main role of this switch, is to
route any single particle from any of its external nodes to any
other simply by single-step control of the central qubits. Were
it not for the existence of a Hadamard matrix in 4 dimensions,
this interesting property could not exist. In fact if one uses any
other matrix instead of the Hadamard, then one looses one or the
other nice property of the switch. For example if one uses a
complex Fourier transform as in \cite{Pemberton}, then multiple
control is needed on the external nodes, and the waves should
traverse the elementary structures several times until by
suitable control of external nodes, they interfere to form a
particle state at a specific output node. On the other hand if
one uses real orthogonal matrices, then the couplings will no
longer be uniform. It would be interesting to see if the scheme
proposed in this letter can be realized experimentally, in solid
state systems, for example in arrays of Joshephson junctions
\cite{Paternostro}. It would also be interesting to see if the
system introduced in this letter can be used as a spin network
quantum computer along the lines set in \cite{KayGeo}.\\

\textbf{Acknowledgement} The authors thank Alastair Kay for his
valuable comments on an earlier version of this letter. M. A. and
V. K. would like to thank L. Memarzadeh for very valuable
discussions.

\end{document}